\documentclass[a4paper,11pt]{article}
\usepackage{pos}

\usepackage[english]{babel}  
\usepackage[utf8]{inputenc} 
\usepackage{hyperref}    
\usepackage[separate-uncertainty=true,multi-part-units=single]{siunitx}
\usepackage{subcaption}
\usepackage{diagbox}
\usepackage{makecell}
\usepackage{verbatim} 
\usepackage{natbib}

\title{Study of the Uncertainties of the Galactic Radio Background as a Calibration Source for Radio Arrays}
\ShortTitle{Uncertainties of the Galactic Radio Background}

\author*[a,b]{Max Büsken}
\author[c]{Tomáš Fodran}
\author[d,e]{Tim Huege}
\affiliation[a]{Institute for Experimental Particle Physics (ETP), Karlsruhe Institute of Technology (KIT),\\ Hermann-von-Helmholtz-Platz 1, 76344 Eggenstein-Leopoldshafen, Germany}
\affiliation[b]{Instituto de Tecnologías en Detección y Astropartículas (CNEA, CONICET, UNSAM),\\ Av. General Paz 1555 (B1630KNA), San Martín, Buenos Aires, Argentina}
\affiliation[c]{Department of Astrophysics/IMAPP, Radboud University, PO Box 9010, 6500 GL, The Netherlands}
\affiliation[d]{Institute for Astroparticle Physics (IAP), Karlsruhe Institute of Technology (KIT),\\ Hermann-von-Helmholtz-Platz 1, 76344 Eggenstein-Leopoldshafen, Germany}
\affiliation[e]{Astrophysical Institute, Vrije Universiteit Brussels,\\ Pleinlaan 2, 1050 Brussels, Belgium}

\emailAdd{max.buesken@kit.edu}
\emailAdd{t.fodran@science.ru.nl}
\emailAdd{tim.huege@kit.edu}

\abstract{The indirect detection of cosmic rays via the radio signal of extensive air showers is gaining a lot of ground. Many new arrays of radio antennas are under construction or in the phase of development. Calibrating these arrays is important for the reconstruction of observed events and for the comparability between observatories. Using reference antennas in calibration campaigns is not ideal because of uncertainties on their signal output strength that are large or difficult to assess. In a different approach the arrays can be calibrated against the Galactic radio emission as the dominant source of background.
This so-called Galactic Calibration relies on predictions of the diffuse Galactic radio emission, for which models are publicly available. We present a comparison of these models in the frequency range from 10 to 408 MHz in order to estimate the systematic uncertainties on the strength of the Galactic background. We do this comparison on a global level as well as adapted for selected radio arrays and discuss implications for applying the Galactic calibration method. Furthermore we study the influence of the quiet Sun as an additional source of radio emission in the sky.}

\FullConference{%
  9th International Workshop on Acoustic and Radio EeV Neutrino Detection Activities - ARENA2022\\
  7-10 June 2022\\
  Santiago de Compostela, Spain}

\begin{document}
\maketitle

\section{Introduction}
\label{sec:1_Introduction}

The radio detection technique in astroparticle physics -- in particular, the detection of ultra-high-energy cosmic rays -- has become mature in recent years \cite{Huege_2016, Schroeder_2017}. Observing MHz radio signals produced in cosmic-ray induced extensive air showers allows for assessing key observables like the depth of shower maximum and the energy of the shower-initiating particle. Accurate measurements of these observables require an accurate calibration of the antenna arrays. An emerging calibration method utilizes the low-frequency Galactic radio emission as it provides the dominant background signal in most arrays \cite{Mulrey_2019, Fodran_2021}. By comparison of background data and predictions of the Galactic signal strength with an understanding of the detector response, an absolute calibration can be made, in principle, at any time and is much easier for large arrays than a calibration campaign with an external reference source.\\
To evaluate the accuracy of the Galactic calibration, the systematic uncertainties on the prediction of the Galactic emission have to be known. Therefore, we present here a comparison of sky models that generate such predictions and estimate these systematic uncertainties for frequencies between \SI{10}{MHz} and \SI{408}{MHz}. After performing the comparison on a global scale, we adjust it to the cases of a few selected radio antenna arrays and determine their individual uncertainties regarding the prediction of the Galactic emission from their local skies. Furthermore, we investigate the influence of the quiet Sun, which is another source of low-frequency radio emission in the sky.

\section{Reference maps and models for predicting the radio sky}
\label{sec:2_Models}
We compare four publicly available models that predict the galactic radio emission and generate sky maps of the brightness temperature at a given frequency: LFmap \cite{LFmap}, the Global Sky Model (GSM) in its original version from 2008 \cite{de_Oliveira_Costa_2008} and in its improved version from 2016 \cite{Zheng_2016}, and the Low Frequency Sky Model (LFSM) \cite{Dowell_2017}. The models take reference maps from radio-astronomical surveys and interpolate between them. The reference maps used by the models for frequencies up to \SI{408}{MHz} and the uncertainties on their temperature scales as determined or estimated by the respective authors are listed in Tab.\ \ref{tab:reference_maps}. The models partly use the same reference maps. Uncertainties on the observed brightness temperatures are within 20\% for all of them. 

\begin{table}
	\centering
	\vspace*{2.0mm}
	\bgroup
	\def\arraystretch{1.2}
	\resizebox{\textwidth}{!}{
	\begin{tabular}{r||r|r|r|r|r|r|r}
	\makecell{map \\ No.} & \makecell{frequency \\ $\nu$/MHz} & \makecell{covered \\ declination} & \makecell{relative \\ scale \\ uncert.\ $\sigma_k$/\%} & \makecell{zero-level \\ error $\sigma_{T_0}$/K} & \makecell{zero-level \\ error norm. \\ to average/\%} & used in & Ref.                       \\ \hline \hline
      1  & 10 & $\SI{-6}{\degree} < \delta < \SI{74}{\degree}$ & 9* & $2{\times}10^4$ & 7.0 & 1,2,3 & \cite{Caswell_1976}             \\ \hline
      2  & 22 & $\SI{-28}{\degree} < \delta < \SI{80}{\degree}$ & 16 & $5{\times}10^3$ & 11.5 & 1,2,3,4 & \cite{Roger_1999}               \\ \hline
      3  & 40 & $\SI{-40}{\degree} < \delta < \SI{90}{\degree}$ & 20 & 10 & 0.1 & 3 & \cite{Dowell_2017}              \\ \hline
      4  & 45 & $\SI{-90}{\degree} < \delta < \SI{65}{\degree}$ & 10/15 & 125$^{\dagger}$ & 1.5 & 1,2,3,(4) & \cite{Alvarez_1997, Maeda_1999} \\ \hline
      5  & 50 & $\SI{-40}{\degree} < \delta < \SI{90}{\degree}$ & 20 & 10 & 0.2 & 3 & \cite{Dowell_2017}              \\ \hline
      6  & 60 & $\SI{-40}{\degree} < \delta < \SI{90}{\degree}$ & 20 & 10 & 0.2 & 3 & \cite{Dowell_2017}              \\ \hline
      7  & 70 & $\SI{-40}{\degree} < \delta < \SI{90}{\degree}$ & 20 & 10 & 0.3 & 3 & \cite{Dowell_2017}              \\ \hline
      8  & 80 & $\SI{-40}{\degree} < \delta < \SI{90}{\degree}$ & 20 & 10 & 0.5 & 3 & \cite{Dowell_2017}              \\ \hline
      9  & 85 & $\SI{-25}{\degree} < \delta < \SI{25}{\degree}$ & 7 & 120 & 6.7 & 2 & \cite{Landecker_1970}           \\ \hline
      10  & 150 & $\SI{-25}{\degree} < \delta < \SI{25}{\degree}$ & 5 & 40 & 9.2 & 2 & \cite{Landecker_1970}           \\ \hline
      11  & 178 & $\SI{-5}{\degree} < \delta < \SI{88}{\degree}$ & 10 & 15 & 5.3 & (1,2,3) & \cite{Turtle_1962}           \\ \hline
      12.a  & 408 (1982) & $\SI{-90}{\degree} < \delta < \SI{90}{\degree}$ & 10/5 & 3 & 8.8 & 1 & \cite{Haslam_1982}              \\ \hline
      12.b  & 408 (2003) & $\SI{-90}{\degree} < \delta < \SI{90}{\degree}$ & 10/5 & 3 & 8.8 & 4 & \cite{Platania_2003}            \\ \hline
      12.c  & 408 (2015) & $\SI{-90}{\degree} < \delta < \SI{90}{\degree}$ & 10/5 & 3 & 8.8 & 2,3 & \cite{Remazeilles_2015}         \\ 
	\end{tabular}}
	\egroup
	\caption{\label{tab:reference_maps} Overview of all reference maps used by the four sky models for frequencies up to \SI{408}{MHz}. The respective authors give the uncertainties of the surveys by considering a linear relation \mbox{$T_\mathrm{true} = k T_\mathrm{obs} + T_0$} between the true temperature $T_\mathrm{true}$ and the observed temperature $T_\mathrm{obs}$. We list the quoted uncertainties on $k$ and $T_0$ (the latter as absolute values and normalized to the average temperature of the sky at the respective frequency). Values of $sigma_k$ with a (*) are estimated by taking half of the smallest contour interval of that map and dividing it by its minimum brightness temperature, because the authors gave no estimate. For zero-level errors with a $(^{\dagger})$ half of the smallest contour interval of the map is taken as the estimate.
    The usage of the reference maps in the individual models is denoted by the following notation: 1 = GSM, 2 = GSM16, 3 = LFSM, 4 = LFmap. Numbers in brackets indicate that information from the reference map was only used indirectly in the respective model.}
\end{table}

The interpolation in the sky models is realized in different ways. In LFmap, spectral scaling with a power law is applied to a single reference map from a survey at \SI{408}{MHz} when evaluating any other frequency. The spectral indices for scaling are taken from measurements in the respective frequency range. In contrast, the other three models use principal component analysis to build a model for predicting the radio sky. More details on the models can be found in Refs.\ \cite{LFmap, de_Oliveira_Costa_2008, Zheng_2016, Dowell_2017}.

\begin{figure}
  \begin{subfigure}[c]{0.48\textwidth}
    \center
    \includegraphics[width=1.\textwidth]{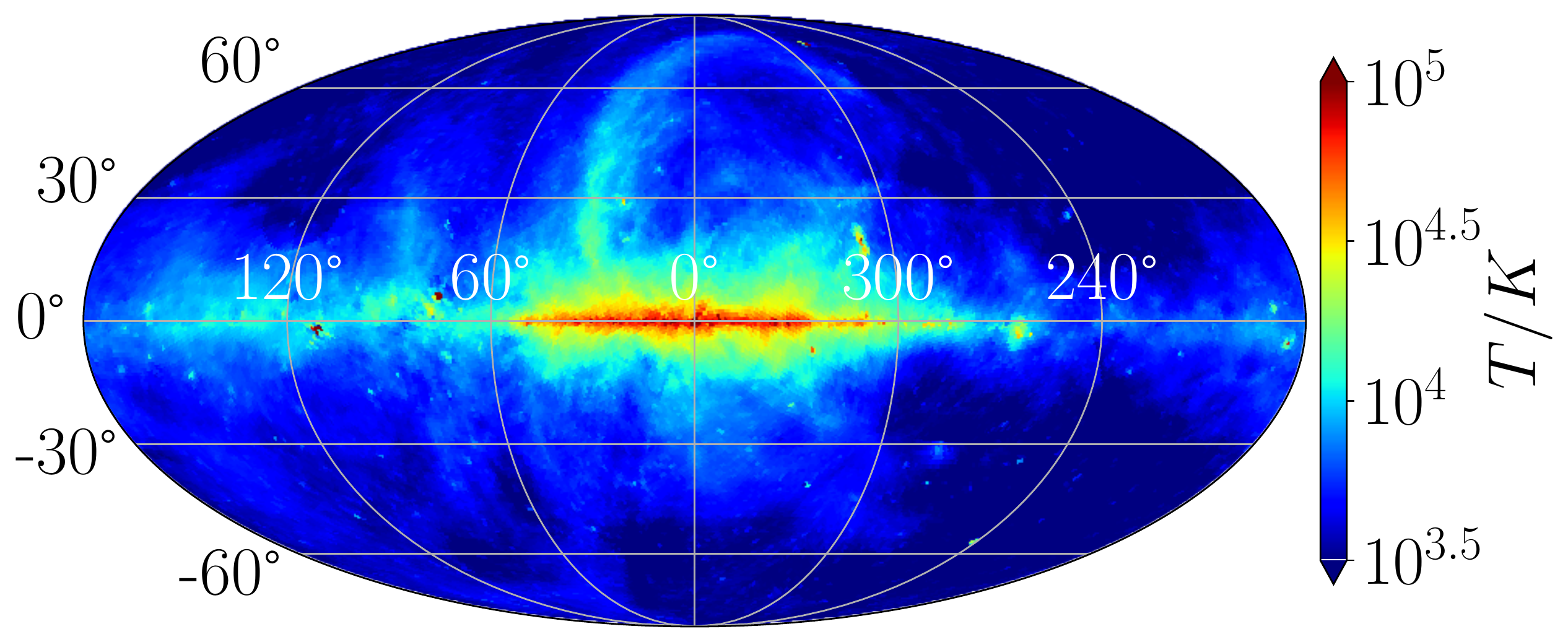}
    \subcaption{LFmap}
  \end{subfigure}\hfill
  \begin{subfigure}[c]{0.48\textwidth}
    \center
    \includegraphics[width=1.\textwidth]{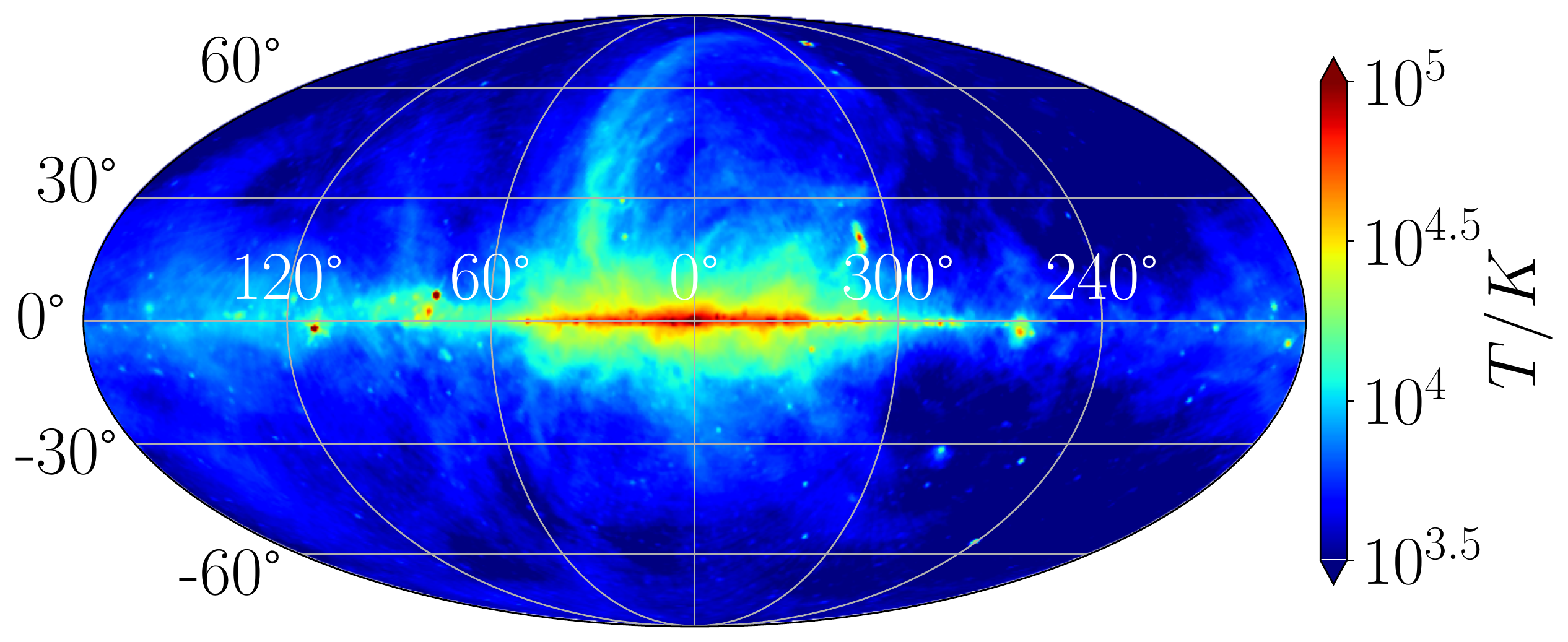}
    \subcaption{GSM}
  \end{subfigure}
  \vspace{5pt}
  \begin{subfigure}[c]{0.48\textwidth}
    \center
    \includegraphics[width=1.\textwidth]{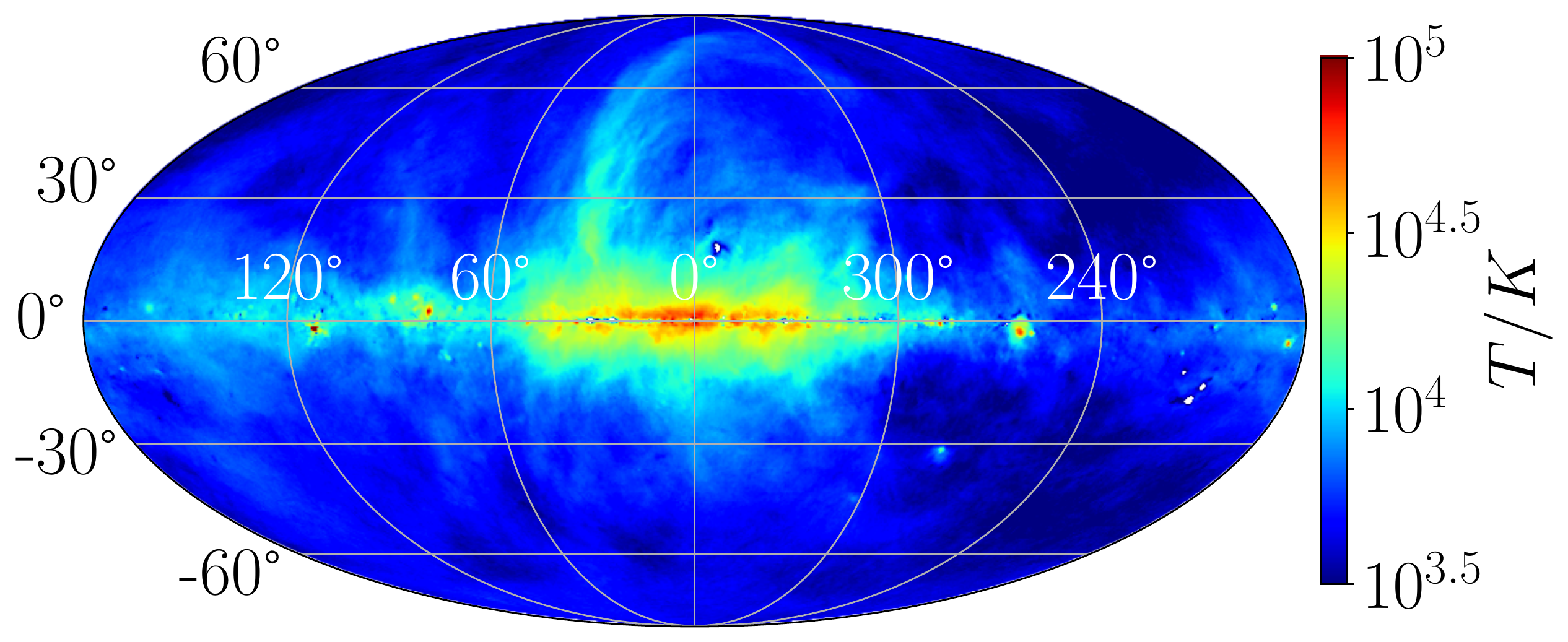}
    \subcaption{GSM16}
  \end{subfigure}\hfill
  \begin{subfigure}[c]{0.48\textwidth}
    \center
    \includegraphics[width=1.\textwidth]{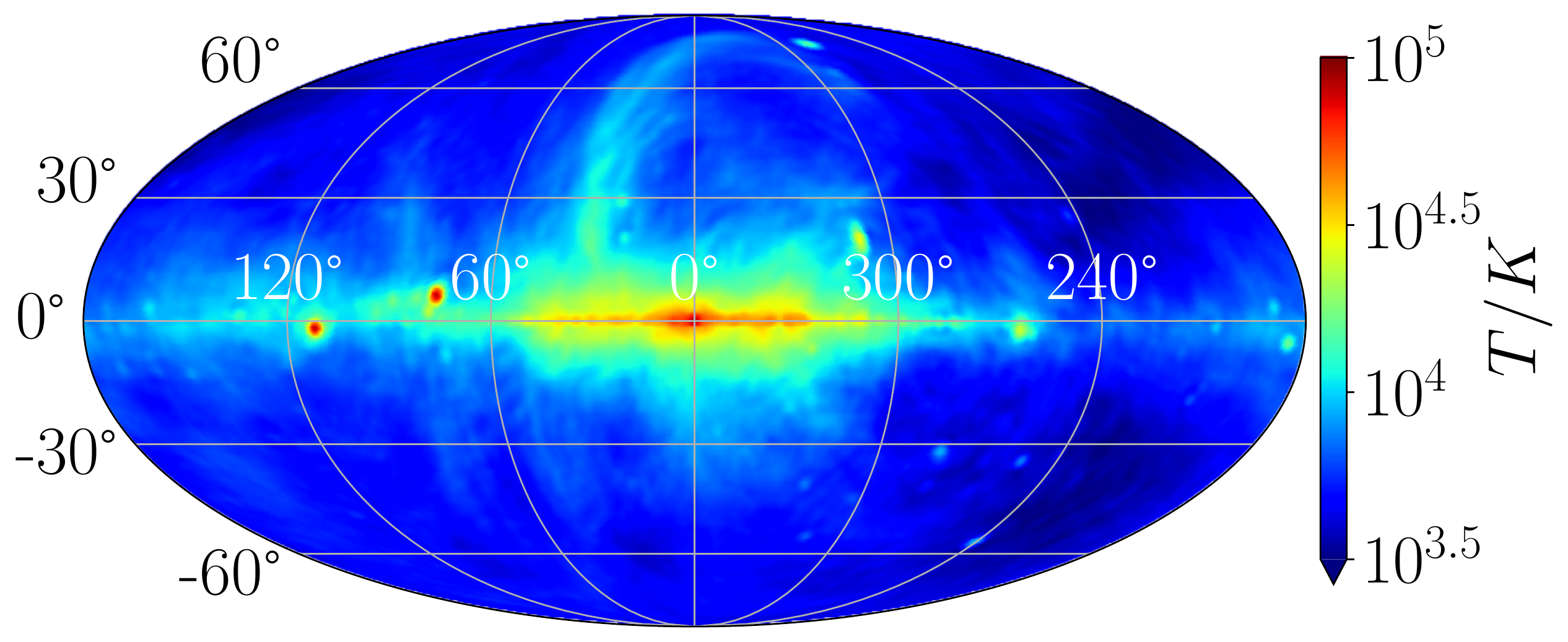}
    \subcaption{LFSM}
  \end{subfigure}
  \caption{Example sky maps at \SI{50}{MHz} produced with each of the four models in galactic coordinates.}
  \label{fig:Example maps}
\end{figure}

As an example, sky maps at \SI{50}{MHz} are shown in Fig.\ \ref{fig:Example maps}, which are generated with the four interpolation models. Differences between the outputs of the models regarding spatial structures and the temperature scales are qualitatively visible. 

\section{Comparison of the sky models}
\label{sec:4_Comparison}
We aim to quantify the differences between the output of the sky models in order to estimate the systematic uncertainty for predicting the radio sky. Therefore, we chose the average sky temperature as a comparative variable which we define as

\begin{equation}
\label{eq:Tsky_average}
\bar{T}(\nu) = \frac{1}{4\pi} \int_{-\pi}^{\pi} \mathrm{d} \ell \int_{\frac{-\pi}{2}}^{\frac{\pi}{2}} \mathrm{d} b \cos{(b)}\; T(\nu; \ell, b).
\end{equation}

Here, $\ell$ and $b$ are the Galactic spherical coordinates and $T(\nu; \ell, b)$ is the brightness temperature at a specific frequency $\nu$ and location in the sky from the output of one of the sky models. We use the Python package PyGDSM \cite{PyGDSM} to obtain the output of the LFSM as well as the two versions of the GSM in the HEALPix \cite{healpix} format. LFmap is not included in PyGDSM and uses a different output format, which we convert to HEALPix and separately feed into the analysis.

\begin{figure}
	\centering
	\includegraphics[width=1.0\textwidth]{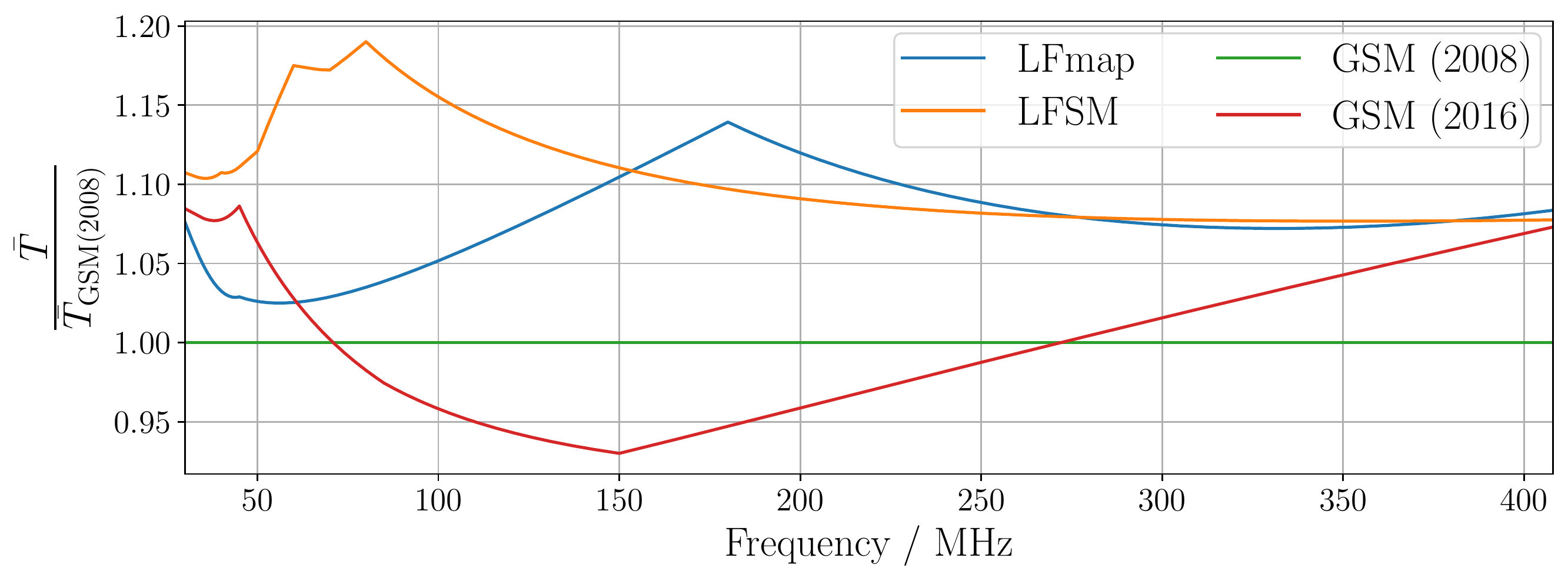}
	\caption{Average sky temperature from the output of the models plotted against the frequency after normalization to the results of the GSM (2008).}
	\label{fig:All_models_normed_log}
\end{figure}

In the considered frequency range, the average sky temperature, in first order, drops exponentially as a function of the frequency. Second-order differences between the models are shown in Fig.\ \ref{fig:All_models_normed_log}, where we plot the average sky temperature obtained from each of the four sky models as a function of the frequency and normalize the curves to the one from the original GSM. Maximum deviations between the models increase when going from \SI{30}{MHz} to higher frequencies and decrease again towards \SI{400}{MHz}, while staying within 20\% overall.

We further quantify how any two of the models agree with each other by evaluating the quantity $r_{\text{m}_1,\text{m}_2}$, which we define as

\begin{equation}
\label{eq:relative_diff}
r_{\text{m}_1,\text{m}_2} = 2 \frac{\int_{\SI{30}{MHz}}^{\SI{408}{MHz}} [ \bar{T}_{\text{m}_1}(\nu) - \bar{T}_{\text{m}_2}(\nu) ] \mathrm{d} \nu }{\int_{\SI{30}{MHz}}^{\SI{408}{MHz}} [ \bar{T}_{\text{m}_1}(\nu) + \bar{T}_{\text{m}_2}(\nu) ] \mathrm{d} \nu }.
\end{equation}

The results of this evaluation are listed in Tab.\ \ref{tab:model_comparison}. Over the whole frequency range, $r_{\text{m}_1,\text{m}_2}$ is smaller than 12\%, which can therefore be considered a global measure for the level of agreement between the sky models. In this evaluation, LFSM stands out against the other models, as it shows the largest deviation when compared to the other three models.

\begin{table}
\centering
\def\arraystretch{1.2}
\begin{tabular}{l||r|r|r|r}
\diagbox{$\text{m}_1$}{$\text{m}_2$} & LFmap & GSM & GSM16 & LFSM \\ \hline \hline
LFmap   &   -    	  &  $4.6\%$    &   $0.7\%$     &   $-7.3\%$ \\ \hline
GSM     &   $-4.6\%$    &   -  	  &   $-3.8\%$    &   $-11.9\%$ \\ \hline
GSM16   &   $-0.7\%$    &  $3.8\%$    &    -   		&   $-8.1\%$ \\ \hline
LFSM    &   $7.3\%$     &  $11.9\%$   &   $8.1\%$    &   - \\
\end{tabular}
\caption{\label{tab:model_comparison} Resulting $r_{\text{m}_1,\text{m}_2}$ tabulated for each model combination m\textsubscript{1}, m\textsubscript{2}.}
\end{table}

\paragraph{Evaluation for astroparticle radio arrays}\text{ }

In the next step, we switch from a global point of view to the specific cases of selected radio arrays. The chosen arrays, which are -- amongst other things -- used in studies of astroparticle physics, are \mbox{RNO-G} (surface antennas) \cite{RNO-G_Design}, LOFAR \cite{LOFAR_Design}, GRAND \cite{GRAND_Design}, SKA-low \cite{SKA-low_Design}, the Pierre Auger Observatory (AERA \cite{AERA} and the AugerPrime Radio Detector \cite{Pont_2019}), and the radio antennas of the IceCube surface array \cite{IceCube_SurfaceArray_Development}.

We adapt the evaluation of the sky model comparison in a simple way to the specifications of each radio array. First, we convert the output maps of the sky models from Galactic to local coordinates (with the two angles azimuth $\alpha$ and altitude $a$, the elevation angle above horizontal) with the observer at the coordinates of the experiment (geographical latitude $\ell_\text{exp}$) and evaluate the average local sky temperature as

\begin{equation}
\bar{T}_{\text{local}}(\nu, \ell_\text{exp}) = \frac{1}{2\pi} \int_{\SI{0}{h}}^{\SI{24}{h}} \mathrm{d} t_{\text{LST}} \int_{\SI{15}{\degree}}^{\SI{90}{\degree}} \cos{(a)} \mathrm{d} a \int_{\frac{-\pi}{2}}^{\frac{\pi}{2}} T(\nu, \ell_\text{exp}, t_{\text{LST}}; a, \alpha) \mathrm{d} \alpha .
\end{equation}

Here, we average the brightness temperature of the local sky $T(\nu, \ell_\text{exp}, t_{\text{LST}}; a, \alpha)$ over the full \SI{24}{h} of local sidereal time (LST) where we limit the contributing sky to everything above \SI{15}{\degree} to mimic the typical directional sensitivity of an antenna.

Next, we adapt the frequency integral to the nominal frequency band of each array with the boundaries $\nu_\text{exp, lower}$ and $\nu_\text{exp, upper}$. With that, we calculate

\begin{equation}
\mathcal{T} (\ell_\text{exp}) = \int_{\nu_\text{exp, lower}}^{\nu_\text{exp, upper}} \bar{T}_{\text{local}}(\nu, \ell_\text{exp}) \; \mathrm{d} \nu .
\end{equation}

Finally, we define $r_{\text{exp; m}_1\text{, m}_2}$ to again measure the level of agreement between any two sky models for a given experiment,

\begin{equation}
r_{\text{exp; m}_1\text{, m}_2} = 2 \frac{ \mathcal{T}_{\text{m}_1} (\ell_\text{exp}) - \mathcal{T}_{\text{m}_2} (\ell_\text{exp}) }{ \mathcal{T}_{\text{m}_1} (\ell_\text{exp}) + \mathcal{T}_{\text{m}_2} (\ell_\text{exp}) }.
\end{equation}

The value of $r_{\text{exp; m}_1\text{, m}_2}$ is listed for all arrays in Tab.\ \ref{tab:experiments_comparison}, once when including all four sky models and once when omitting LFSM. The maximum deviation is between 10\% and 19\%, depending on the experiment. These numbers reduce significantly to between 2\% and 14\% when excluding LFSM, showing a discrepancy between the output of this model and the other three.

\begin{table}
\centering
\def\arraystretch{1.2}
\resizebox{\textwidth}{!}{
\begin{tabular}{r||r|r!{\vrule width 2pt}r|r!{\vrule width 2pt}r|r}
experiment & \makecell{geographic \\ latitude} & \makecell{frequency \\ band / MHz} & \makecell{maximum \\ relative \\ deviation \\ (incl.\ LFSM)} & \makecell{corresponding \\ sky models} & \makecell{maximum \\ relative \\ deviation \\ (excl.\ LFSM)} & \makecell{corresponding \\ sky models} \\ \hline \hline
RNO-G \cite{RNO-G_Design}   & \SI{ 72.58}{\degree}  & \SIrange{100}{408}{} & 17.1\% & LFSM/GSM16   & 9.6\% & LFmap/GSM16 \\ \hline
LOFAR low \cite{LOFAR_Design}   & \SI{ 52.91}{\degree}  & \SIrange{30}{80}{}  & 13.3\% & LFSM/GSM   & 8.9\% & GSM/GSM16 \\ \hline
LOFAR high \cite{Nelles_2015}   & \SI{ 52.91}{\degree}  & \SIrange{110}{190}{}  & 18.4\% & LFSM/GSM16   & 12.8\% & LFmap/GSM16 \\ \hline
GRAND \cite{GRAND_Design}   & \SI{ 42.93}{\degree}  & \SIrange{50}{200}{}  & 16.3\% & LFSM/GSM   & 2.0\% & LFmap/GSM \\ \hline
SKA-low \cite{SKA-low_Design} & \SI{ -26.70}{\degree} & \SIrange{50}{350}{}  & 13.6\% & LFSM/GSM16 & 7.5\% & LFmap/GSM16 \\ \hline
Auger \cite{AugerPrime_Radio}   & \SI{ -35.21}{\degree} & \SIrange{30}{80}{}   & 10.5\% & LFSM/GSM   & 4.8\% & LFmap/GSM \\ \hline
IceCube \cite{IceCube_SurfaceArray_Development} & \SI{ -90.0}{\degree}  & \SIrange{70}{350}{} & 18.6\% & LFSM/GSM16 & 13.6\% & LFmap/GSM16 \\ \hline
\end{tabular}}
\caption{\label{tab:experiments_comparison} Levels of agreement of the sky models for each of the selected radio antenna arrays evaluated from the comparisons based on the local skies and nominal frequency bands of the respective array. The frequency band of RNO-G is capped at \SI{408}{MHz} due to the limitations of LFmap and LFSM, although the surface antennas will also cover higher frequencies.}
\end{table}

\section{Influence of quiet Sun}
\label{sec:5_AdditionalSources}

\begin{figure}
	\centering
	\includegraphics[width=1.0\textwidth]{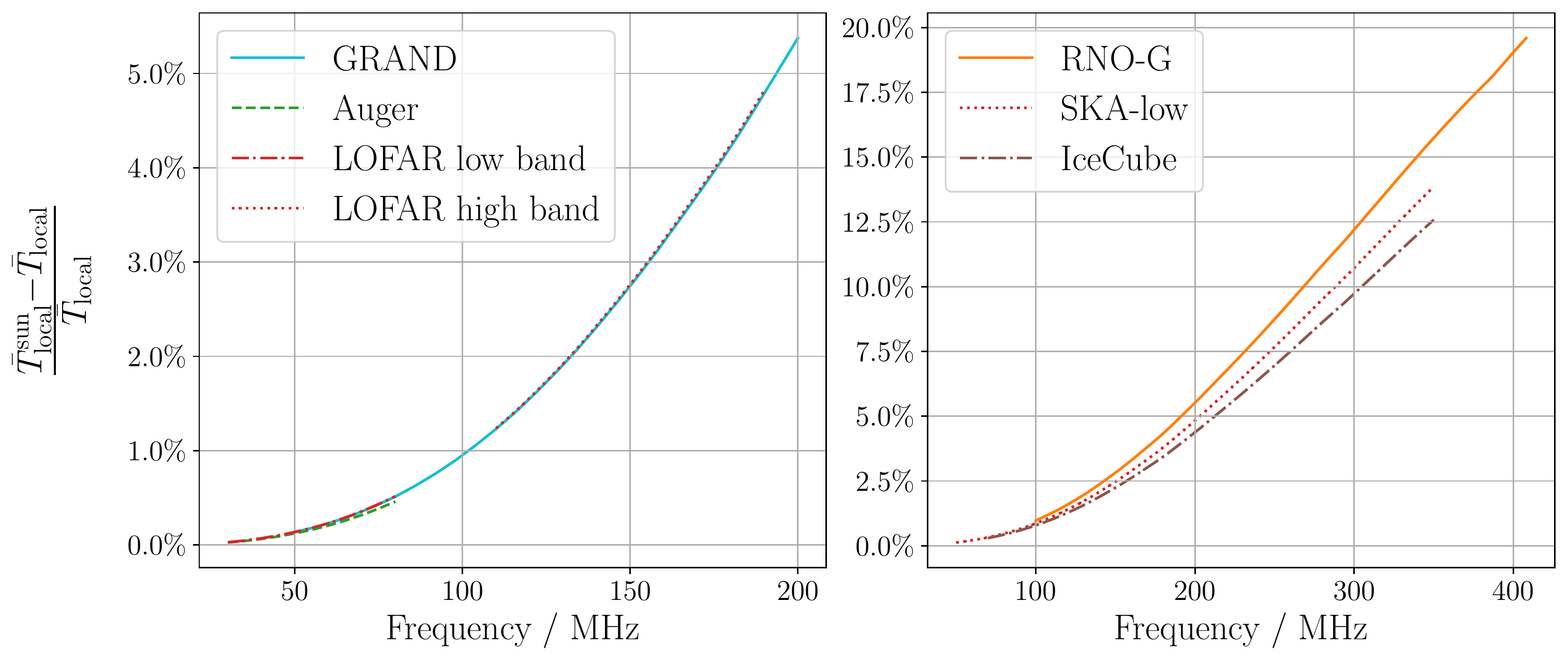}
	\caption{Relative difference in the average temperature of the local sky induced by the quiet Sun plotted for the selected radio arrays. The lines represent the average results from using the four sky models to produce the maps, while the colored bands show any model's maximum and minimum contribution.}
	\label{fig:All_models_local_sun}
\end{figure}

In otherwise radio-quiet regions, the Galaxy is the dominant background contribution for radio arrays. However, subdominant contributions can alter the accuracy of the Galactic calibration, e.g., radio emission from the Sun \cite{Kraus_RadioAstronomy}. Therefore, we study the influence of the quiet Sun, i.e., without considering solar flares, by its relative strength compared to the Galaxy. We place a circularly shaped spot of homogeneous brightness temperature on the output maps of the models for the local skies of the selected radio arrays. The solar brightness temperature as a function of the frequency is taken from Ref.\ \cite{Zhang_2022}. We calculate the relative difference of the average sky brightness temperature with and without the superimposed Sun. This relative difference is shown as a function of the frequency for all experiments in Fig.\ \ref{fig:All_models_local_sun}.

In general, the relative solar contribution increases with frequency. It is insignificant for arrays operating at lower frequencies, like the Auger radio detectors or LOFAR in the low-band configuration, while the contribution rises to around 20\% at \SI{400}{MHz}, which is relevant for RNO-G, SKA-low, and IceCube.

\section{Discussion and conclusion}
\label{sec:6_Discussion}

The comparison of four radio sky models conducted in this work allows for an estimation of the uncertainty on how accurately the radio sky can be predicted at low frequencies. This uncertainty is relevant for applying the Galactic calibration to radio antenna arrays. In literature research on the reference surveys used in making the sky models, we find that uncertainties on the temperature scales can be as large as 20\%, which propagates into the accuracy of the models. We compare the output of the models via the average temperature of the entire sky and find a level of agreement of 12\%. Furthermore, we adapt the comparison to selected radio antenna arrays by observing the local sky instead of the entire sky and using the operational frequency band of the arrays for cosmic-ray detection. The comparison evaluation yields different results for the variety of arrays, ranging from 10\% for the Auger radio detectors up to 19\% for the radio array of IceCube.\\
In addition, we study the contribution of the quiet Sun to the radio sky background and find it to evolve from an insignificant influence at the lowest considered frequencies to around 20\% at \SI{400}{MHz}. Consequently, the quiet Sun can be neglected for arrays at the lowest frequencies, like Auger and LOFAR in the low-band mode, while it should definitely be considered for the higher-frequency arrays, like RNO-G, SKA-low, and IceCube. The latter may need to take measures in that regard when applying the Galactic calibration, e.g., by restricting to night times. 
The relative uncertainties on the brightness temperature determined in this work apply to the received power in the antenna, which scales with the square of the electric-field amplitudes of radio signals from detected particles and thus with the square of the energy scale of these particles. Therefore, uncertainties on the energy scale are about half of the uncertainties on the temperature. In this light, the Galactic calibration turns out to be on par with or better than typical calibration methods using an external reference source \cite{Aab_2017, Mulrey_2019}. In the future, additional sky surveys and improved models for predicting the Galactic radio emission will increase the accuracy and reduce the systematic uncertainties associated with the strength of the Galactic radio background. This outlook will make the Galactic calibration the future standard for the calibration of radio arrays in astroparticle physics.\\

\setlength{\bibsep}{0pt plus 0.3ex}
\bibliographystyle{MyJHEP}
\bibliography{refsm}


\end{document}